# Reduced Neutrino Opacities and the SN 1987A Signal

Wolfgang Keil and H.-Thomas Janka
Max-Planck-Institut für Astrophysik
85740 Garching, Germany

and

Georg Raffelt
Max-Planck-Institut für Physik
Föhringer Ring 6, 80805 München, Germany

October 5, 1994

**Abstract**

Axial-vector interactions of neutrinos and axions with a medium are based on structure functions which cannot be calculated in simple perturbative terms for nuclear densities. We use the SN 1987A neutrino signal duration to estimate the range of allowed neutrino opacities of a supernova (SN) core. We perform numerical simulations of the evolution from the hot, lepton-rich post-collapse stage to the cold, deleptonized neutron star and compare the predicted neutrino signal characteristics with the SN 1987A data. Besides the neutrino opacities we vary the mass, temperature profile, and equation of state of the initial model. Within plausible limits of these quantities the neutrino opacity cannot be much smaller than its "standard" value. This limits the width of the spin-density structure function to much smaller values than indicated by a perturbative calculation, implying that bremsstrahlung processes such as axion emission probably saturate at around 10% nuclear density. A substantial suppression at yet larger densities that might have been expected does not seem to occur.

# 1 Introduction

Aside from gravitational waves neutrinos provide the only known possibility to obtain direct information about the physical processes in matter at the extreme conditions of density and temperature that occur at the center of type-II supernova (SN) explosions, in colliding or coalescing binary neutron stars, or in collapsing white dwarfs. A type-II SN explosion marks the violent final stage in the life of a massive star with a main sequence mass $8\,M_\odot \lesssim M \lesssim 30\,M_\odot$. It is associated with the disruption of the star and an outburst of light with a total energy of about $10^{49}$ erg. Most of the stellar mantle and envelope is ejected with a kinetic energy of the order of $10^{51}$ erg, forming the ultimately diffuse SN remnant. The progenitor star's core with a mass of $1$–$2\,M_\odot$ is thought to form a compact object, i.e. a neutron star or possibly a black hole.

Core collapse events and merging binary neutron stars are powerful sources of neutrinos which are far more efficient at carrying away energy from the dense inner regions than photons for which the opacity is about 20 orders of magnitude larger. 99% of the gravitational binding energy (a few $10^{53}$ erg) of the neutron star forming in a SN are emitted in neutrinos and antineutrinos of all flavors. They carry away lepton number and energy, thus driving the evolution from the hot, lepton-rich collapsed stellar core to the cold, neutronized remnant. In addition, recent progress in our understanding of SN physics indicates that neutrinos play an active and crucial role for different processes during type-II SN explosions. Neutrino energy deposition in the layers between the newly formed neutron star and the SN shock front is thought to yield the energy for ejecting the stellar mantle and envelope [1]. This energy deposition creates a "hot bubble" outside of the protoneutron star which has comparatively low density, but quite high temperatures [2]. Also, the transfer of energy from neutrinos to the material near the surface of the protoneutron star causes a continuous flow of matter into the hot-bubble region. This neutrino-driven wind may well be the long-sought site for the r-process nucleosynthesis [3].

The measured SN 1987A neutrino signal at the Kamiokande II and IMB water Cherenkov detectors [4] crudely confirms theoretical expectations. Notably, the duration of the signal of several seconds precludes efficient cooling via "invisible channels" such as the emission of right-handed neutrinos or axions. Therefore, the SN 1987A signal has been used to set a large number of constraints on a variety of novel particle physics models and phenomena [5]. Using SN 1987A and perhaps a future galactic SN as a "particle physics laboratory" along these lines, as well as a proper understanding of the SN explosion mechanism and r-process nucleosynthesis all require a reliable calculation of the "neutrino light curve" from a core collapse SN.

To this end the transport of neutrinos through the stellar medium has to be simulated in numerical models of SN explosions and of the protoneutron star evolution, ideally by solving the Boltzmann transport equation. In practice, an approximate treatment based on the corresponding moment equations is usually employed. A necessary input are the dynamical structure functions of the stellar medium which account for the interactions with the neutrinos. For instance, the neutral-current scattering $\nu(k_1) \to \nu(k_2)$ is completely described by a function $W(k_1, k_2)$ which gives the probability for a transition from a neutrino with four momentum $k_1$ to one with $k_2$. For most practical applications this function was calculated on the basis of a simple cross section for the scattering process $\nu(k_1) + N \to N + \nu(k_2)$ on free nucleons $N$. In the nonrelativistic limit nucleon recoils can be neglected; the scattering cross section is then found to be $\sigma = (G_F^2/4\pi)(C_V^2 + 3C_A^2)\varepsilon_1^2$ with the Fermi constant $G_F$, the vector and axial-vector coupling constants[1] $C_V$ and $C_A$, and the initial neutrino energy $\varepsilon_1$. The scattering cross section and the neutrino opacities are thus found to be dominated by axial-vector (spin-dependent) interactions. The same holds true for charged-current reactions of the form $\nu_e + n \leftrightarrow p + e^-$ and $\overline{\nu}_e + p \leftrightarrow n + e^+$.

In a recent paper Raffelt and Seckel [6] have questioned the validity of this approach for the conditions of a SN core. Simply put, they argued that the collisions among nucleons which interact by a spin-dependent force cause the nucleon spins to fluctuate fast relative to other relevant time scales. For neutrino scattering such a time scale is the inverse of the energy transfer $\Delta\varepsilon = \varepsilon_2 - \varepsilon_1$. Neutrinos cannot "resolve" processes which happen at a faster rate and so, if the spin of a given nucleon flips several times during the period $(\Delta\varepsilon)^{-1}$ the neutrino will see an approximately vanishing nucleon spin. For the vector current, a similar effect does not exist because in the nonrelativistic limit there are only small nucleon velocity fluctuations. This difference between the vector and axial-vector interactions is also illustrated by the absence of neutrino pair bremsstrahlung $N + N \to N + N + \overline{\nu} + \nu$ from the vector current, while there is a large contribution from the axial vector. In this case neutrino pairs are emitted by the fluctuating nucleon spins which act as a time-varying source for the neutrino axial-vector current.

In an extreme scenario, the axial-vector cross section could be almost entirely suppressed at high densities, leaving only the vector-current interaction as an opacity source. At first sight this picture appears ruled out by the good agreement between the SN 1987A neutrino observations with standard predictions. In a preliminary study [7] it was found, however, that this agreement might be deceiving in the sense that a modified neutrino spectrum and a modified emission period could conspire such as to leave the observed event numbers and signal durations relatively unchanged. Presently we formalize this study in order to understand the impact of modified (reduced) neutrino opacities as predicted by Raffelt and Seckel [6]. We perform numerical simulations of the neutrino cooling of nascent neutron stars for modified neutrino opacities and compute the expected neutrino signals in the Kamiokande II and IMB detectors. Asking for consistency between these signals and the observed ones we attempt to infer limits on the possible amount by which the effective neutrino cross sections could be reduced in the stellar plasma.

The paper is organized as follows. In Sect. 2 we summarize the considerations in favor of a possible suppression of the axial-vector weak currents at SN conditions. We motivate a simple formula for the suppression effect as a function of density and temperature. A parametric ansatz allows us to switch continuously between full suppression and no suppression of axial-vector currents in our numerical simulations of the neutrino-driven cooling of protoneutron stars (Sect. 3.1). Aside from modified neutrino opacities for neutral-current and/or charged-current processes, we consider models with different neutron star masses, with different versions of the equation of state (EOS) of matter around and above nuclear density, and with different initial temperature profiles. Section 3.2 contains a comparison of expected and observed neutrino signals, and Sect. 4 closes the paper with a discussion and interpretation of our results.

---

[1]In vacuum, the values are $C_V^n = -1$, $C_V^p = 1 - 4\sin^2\Theta_W = 0.07$, $C_A^n \approx -1.15$, and $C_A^p \approx 1.37$. The previously used values $C_A^p = -C_A^n = 1.26$ had to be modified because some of the nucleon spin appears to be carried by strange quarks. (For references see [6].) In a bulk nuclear medium these coefficients are likely suppressed, in analogy to the suppression of the charged current where $C_A$ is suppressed from its vacuum value of 1.26 to about 1.0 in a nuclear medium.

## 2 Neutrino scattering rate

### 2.1 The dynamical spin-structure function

We are mostly concerned with the neutral-current axial-vector scattering rate of neutrinos on the nuclear medium. As a simple model we consider a medium consisting of only one species of nucleons with a number density equivalent to the baryon density $n_B$. The differential scattering cross section per nucleon for the transition $\nu(\varepsilon_1) \to \nu(\varepsilon_2)$ can then be expressed in the form

$$\frac{d\sigma_A}{d\varepsilon_2} = \frac{3C_A^2 G_F^2 \varepsilon_2^2}{4\pi} \frac{S_\sigma(\varepsilon_2 - \varepsilon_1)}{2\pi}, \tag{1}$$

where $G_F$ is the Fermi constant and $C_A$ the neutral-current axial-vector coupling constant with the values for protons and neutrons discussed in the introduction.

In the limit of nonrelativistic nucleons, and ignoring nucleon recoils as well as collective spin motions, one finds that $S_\sigma(\Delta\varepsilon)$ is the spin-density dynamical structure function in the long-wavelength limit (e.g. [6], [8])

$$S_\sigma(\omega) = \frac{1}{3n_B} \lim_{\boldsymbol{k} \to 0} \int_{-\infty}^{+\infty} dt\, e^{i\omega t} \left\langle \boldsymbol{s}(t,\boldsymbol{k}) \cdot \boldsymbol{s}(0,-\boldsymbol{k}) \right\rangle. \tag{2}$$

Here, $s_i = \psi^\dagger \sigma_i \psi$ ($i = 1,2,3$, Pauli matrix $\sigma_i$, nucleon Pauli spinor $\psi$) is the spin-density operator. A calculation of this correlator (and more complicated ones in a mixed medium of several types of constituents) is the main problem at treating neutrino transport correctly.

For a medium of nucleons which do not interact with each other, i.e., in the limit of a very dilute medium one easily finds

$$S_\sigma(\omega) = 2\pi\, \delta(\omega). \tag{3}$$

This form reflects that neutrinos do not change their energy in collisions with isolated "heavy" nucleons (no recoil!). The total axial-vector scattering cross section is then $\sigma_A = 3C_A^2 G_F^2 \varepsilon_1^2/4\pi$.

In the long-wavelength limit the structure function $S_\sigma$ governs all processes which rely on a coupling to the nucleon spin such as the axial-vector $\nu\bar{\nu}$ emission or absorption or the emission or absorption of axions. Axions are hypothetical pseudoscalar particles which couple to nucleons by virtue of the axial-vector interaction Lagrangian $(C/2f_a)\, \overline{\psi}\gamma_\mu\gamma_5\psi\, \partial^\mu a$ where $C$ is a model-dependent dimensionless number of order unity and $f_a$ is the axion decay constant or Peccei-Quinn scale, a quantity with the dimension of an energy. If axions can escape freely from a neutron star so that one may ignore re-absorption and Bose-stimulation effects, the energy-loss rate per unit volume is given by

$$Q_a = \left(\frac{C}{2f_a}\right)^2 \frac{n_B}{4\pi^2} \int_0^\infty d\omega\, \omega^4\, S_\sigma(\omega). \tag{4}$$

In the dilute limit where $S_\sigma(\omega) \to 2\pi\delta(\omega)$ the axion emission rate vanishes.

When nucleon-nucleon interactions are not ignored, the lowest-order contribution to axion emission is bremsstrahlung $N + N \to N + N + a$. An explicit calculation was performed by Brinkmann and Turner [9] in what corresponds to the long-wavelength limit. They modelled the nucleon-nucleon interaction by a one-pion exchange potential with massless pions. Their result can be brought into the form

$$S_\sigma(\omega) = \Gamma_\sigma \, \omega^{-2} \, s(\omega/T) \tag{5}$$

where $T$ is the temperature and $\Gamma_\sigma$ represents something like a nucleon spin rate of change due to collisions. It is found to be [6]

$$\gamma_\sigma \equiv \frac{\Gamma_\sigma}{T} = \frac{4\sqrt{\pi}\,\alpha_\pi^2 n_B}{T^{1/2} m_N^{5/2}} \approx 35 \, \frac{\rho}{\rho_0} \left(\frac{10\,\text{MeV}}{T}\right)^{1/2} \tag{6}$$

where $\alpha_\pi = (f 2m_N/m_\pi)^2/4\pi \approx 17$ with $f \approx 1.05$ is the "pion-nucleon fine structure constant" and $\rho_0 = 3\times 10^{14}\,\text{g cm}^{-3}$ is the nuclear density. The factor $\omega^{-2}$ is easily shown to represent the nonrelativistic propagator of the intermediate nucleon in the bremsstrahlung graph [6]. It can also be shown to follow from a classical treatment of the bremsstrahlung process [10].

The remaining dimensionless function $s(x)$ of the dimensionless energy transfer $x \equiv \omega/T$ represents the nucleon phase space volume in $NN$-collisions. For positive energy transfers (energy given to the neutrino in a collision) it is found to be

$$s(x) = e^{-x} \int_0^\infty dy \, e^{-y} \left(|x|\, y + y^2\right)^{1/2}. \tag{7}$$

It has the asymptotic behavior $s(0) = 1$ and $s(x \gg 1) = e^{-x}(x\pi/4)^{1/2}$. For negative energy transfers one finds that the factor $e^{-x}$ must be replaced by 1. Then

$$S_\sigma(\omega) = S_\sigma(-\omega) \, e^{-\omega/T} \tag{8}$$

as required by detailed balance.

The total cross section based on Eq. (1) together with Eq. (5) diverges because of the $\omega^{-2}$ behavior of the structure function. Raffelt and Seckel [6] proposed to modify this divergence by a Lorentzian cutoff $\omega^{-2} \to (\omega^2 + \Gamma^2/4)^{-1}$ with a suitable energy scale $\Gamma$. Its magnitude is dictated by the normalization requirement

$$\int_{-\infty}^{+\infty} d\omega \, S_\sigma(\omega) = 2\pi \tag{9}$$

which can be shown to follow from the definition of $S_\sigma$ if there are no correlations between the spins of different nucleons in the medium. For $\Gamma_\sigma \ll T$ one finds $\Gamma = \Gamma_\sigma$, i.e., the structure function $S_\sigma(\omega)$ is essentially a Lorentzian of width $\Gamma_\sigma$ which for small $\Gamma_\sigma$ approaches the $\delta$-function already found for a gas of non-interacting nucleons.

The normalization condition illuminates the intuitive fact that the total neutrino scattering rate on nucleons is not increased just because the nucleons interact with each other before and after the interaction with the neutrino. However, these "upstream" or "downstream" collisions do allow for the transfer of energy between nucleons and neutrinos that would not be possible otherwise because of the smallness of nucleon recoils. It is also clear that one cannot conceptually separate the elastic single-nucleon scattering process $\nu + N \to N + \nu$ from the inelastic process $\nu + N + N \to N + N + \nu$ because any nucleon with which a neutrino can interact is known with certainty to have emerged from a previous $NN$-collision, and with certainty will interact again. Therefore, the emerging picture is that $S_\sigma(\omega)$ is a narrow $\delta$-like function in a dilute medium, and that it develops "wings" as the medium becomes denser. The "wings" are probably well determined by the lowest-order bremsstrahlung calculation while the region around $\omega = 0$ is dominated by multiple scattering effects.

## 2.2 $S_\sigma(\omega)$ at high density

As long as $\Gamma_\sigma \ll T$ the picture of neutrino scattering on quasi-free nucleons and the bremsstrahlung version of $S_\sigma(\omega)$ can be reconciled with each other without much problems by the above Lorentzian ansatz for the low-$\omega$ behavior. This "dilute limit" corresponds physically to the picture that a nucleon spin flips only rarely relative to the time scale set by inverse thermal frequencies of the system. In a SN core, however, one is confronted with the opposite limit $\Gamma_\sigma \gg T$, i.e., nominally the nucleon spin flips many times within one oscillation period of, say, a thermal neutrino.

In order to remind the reader of the typical physical conditions present in a young SN core we show in Fig. 1 the profiles of various relevant parameters for our reference case, model S2BH_0 (see Sect. 3), about 1 s after the numerical simulation of the neutrino cooling was started. This stage of the evolution corresponds to the situation in a protoneutron star 1–2 s after its formation. The degeneracy parameter of neutrons is typically between 2 and 4, i.e., neutrons are mildly degenerate. However, for most purposes a value $\eta \approx 3$ marks the borderline between degenerate and nondegenerate conditions whence we think it is justified to use the nondegenerate limit for our exploratory study in this paper. The parameter $\Gamma_\sigma/T$ is found to be typically of order 30, i.e., $\Gamma_\sigma \gg T$.

It is not at all obvious what $S_\sigma(\omega)$ will look like for such conditions. Besides the detailed-balance condition Eq. (8) which must apply on fundamental grounds one may assume that the normalization condition Eq. (9) remains valid (absence of spin-spin correlations), and one may assume that for $\omega \gg \Gamma_\sigma$ the bremsstrahlung result Eq. (6) remains valid. Together, these three conditions imply that $S_\sigma(\omega)$ is a broad and shallow function as opposed to a $\delta$-function in the dilute limit. As a generic example for the shape of $S_\sigma(\omega)$ in the high-density limit we require the asymptotic high-$\omega$ behavior $S_\sigma(\omega) = T^{-1} (\pi/4)^{1/2} \gamma_\sigma x^{-3/2} e^{-x}$ where $x = \omega/T$. Moreover, we speculate that for $\omega \lesssim \Gamma_\sigma$ the structure function is essentially a constant. Then, an ansatz for its shape is

$$\hat{S}_\sigma(x) = \frac{(\pi/4)^{1/2} \gamma_\sigma}{|x|^{3/2} + \beta} \times \begin{cases} e^{-x} & \text{for } x > 0, \\ 1 & \text{for } x < 0. \end{cases} \qquad (10)$$

Here, $\hat{S}_\sigma(x)$ is the dimensionless function $T S_\sigma(Tx)$. For $\gamma_\sigma \gg 1$ the normalization integral is dominated by negative energy transfers whence $\beta = (\pi/27)^{3/2} \gamma_\sigma^3$. This example illustrates the implications of the normalization condition in conjunction with the perturbative calculations which fix the high-$\omega$ asymptotic form.

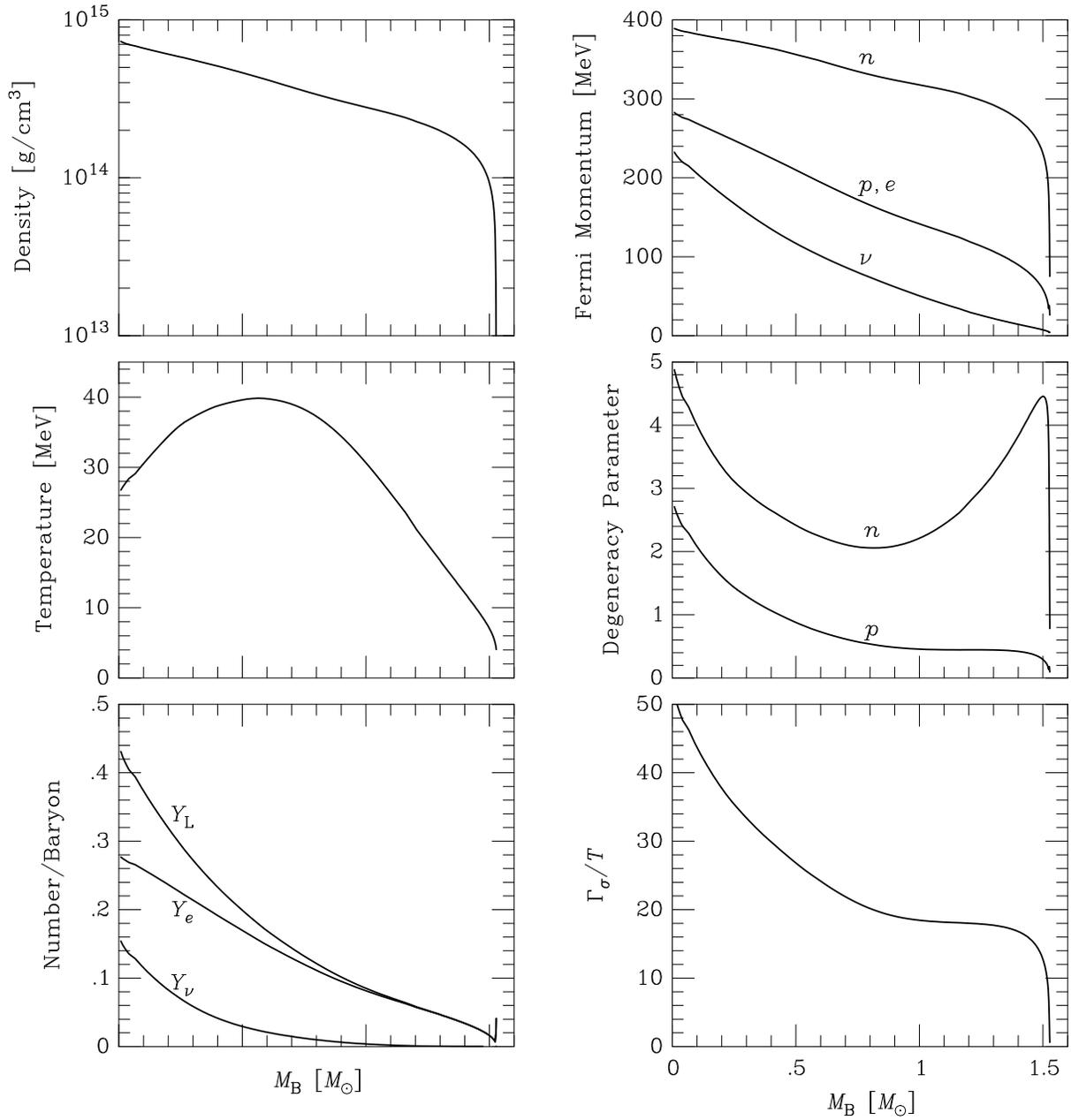

FIG. 1: Physical parameters of our reference model S2BH_0 after about 1 s of neutrino cooling and deleptonization, corresponding to the situation in a protoneutron star about 1–2 s after stellar core collapse. The plots in the left column show profiles of density $\rho$ (top), temperature $T$ (middle), and the number fractions of electron neutrinos, $Y_\nu$, electrons, $Y_e$, and leptons (electrons plus electron neutrinos), $Y_L$ (bottom) as functions of the enclosed baryonic mass $M_B$. The right column of figures displays the Fermi momenta of neutrons, protons, electrons, and neutrinos (top), the corresponding degeneracy parameters of $n$ and $p$ as ratios of the Fermi energies to the temperature (middle), and the parameter $\gamma_\sigma = \Gamma_\sigma/T$ of Eq. (6) (bottom).

The impact of this high-density behavior of $S_\sigma(\omega)$ on the neutral-current neutrino opacity is crudely estimated by the inverse mean free path, averaged over a thermal energy spectrum of the initial neutrino. Moreover, all expressions become much simpler if one replaces the Fermi-Dirac occupation numbers with the Maxwell-Boltzmann expression $e^{-\varepsilon_i/T}$; the resulting error is small for nondegenerate neutrinos. The relevant quantity is then

$$F \equiv \left\langle \frac{\lambda_0}{\lambda} \right\rangle = \frac{1}{48\pi} \int_0^\infty dx_1 \int_0^\infty dx_2\, x_1^2\, x_2^2\, e^{-x_1}\, \hat{S}_\sigma(x_2 - x_1), \tag{11}$$

where $\lambda_0^{-1} = (3C_A^2 G_F^2 n_B/4\pi)\, 24\, T^2$. One energy integration can be done explicitly, leaving one with an integral over the energy transfer alone ($x = \omega/T$)

$$F = \frac{1}{12\pi} \int_0^\infty dx\, (12 + 6x + x^2)\, \hat{S}_\sigma(x). \tag{12}$$

This expression is unity for elastic scattering, i.e., when $\hat{S}_\sigma(x) = 2\pi\delta(x)$ or at least a narrow, normalized function.

For $\gamma_\sigma \gg 1$ one finds with Eq. (10) $F \approx (135\sqrt{3}/2\pi^2)\gamma_\sigma^{-2} \approx 12\,\gamma_\sigma^{-2}$. Therefore, the total neutrino scattering rate drops precipitously for $\gamma_\sigma$ larger than a few. In order to interpolate to the low-density regime where $F = 1$ one may use $F = (1 + \gamma_\sigma^2/12)^{-1}$ where $\gamma_\sigma = \Gamma_\sigma/T$ was given in Eq. (6). With $\gamma_\sigma \approx 20$ in the SN core, the axial-vector neutral-current scattering rate would be entirely suppressed.

While this radical conclusion follows from our three assumptions concerning $S_\sigma$, one or more of them may not apply. We are most suspicious of the assumption that the width of $S_\sigma(\omega)$ is indeed given by $\Gamma_\sigma$ of Eq. (6) as this quantity essentially represents the spin rate of change due to nucleon-nucleon collisions. As colliding nucleons themselves are subject to other collisions, their spins fluctuate as well, leading to an averaging effect which may well reduce $\Gamma_\sigma$ below its perturbative value. It does not seem plausible that the spins will fluctuate much faster than thermal frequencies given by the energy levels of the system. Thus, one may speculate that $S_\sigma$ does not grow much broader once its width is of order $T$. This sort of behavior would imply, for example, that the axion emission rate reaches some maximum level at which it saturates without dropping at yet higher densities. The axial-vector neutrino scattering rate would be suppressed, but perhaps not entirely.

The main goal of our present study is to find out if the neutrino signal of SN 1987A allows us to distinguish between such speculations, and notably, if it allows us to exclude the most radical scenario of a complete suppression of the axial-vector scattering rate.

Keeping in line with these exploratory goals of our study we implement modified neutrino opacities by the prescription $C_A^2 \to FC_A^2$ with

$$F = (1 - a) + \frac{a}{1 + b} \tag{13}$$

where

$$b = \frac{1}{12} \left(\frac{\Gamma_\sigma}{T}\right)^2 \approx \left(\frac{\rho}{3 \times 10^{13}\, \text{g cm}^{-3}}\right)^2 \left(\frac{10\,\text{MeV}}{T}\right). \tag{14}$$

Here, $a = 1$ corresponds to the full suppression effect while $a = 0$ to no suppression at all. For completeness we also consider $a < 0$, implying enhanced cross sections.

While the discussion of Raffelt and Seckel [6] focussed on neutral-current processes one may speculate that similar reduction effects would pertain to charged-current processes. Therefore, we will also investigate the case where the same axial-vector reduction factor applies to both types of reactions.

## 3 Neutrino cooling of protoneutron stars

During the first seconds of its life the newly formed neutron star evolves from a hot, lepton-rich state to a cold, neutronized object, a process driven by the loss of lepton number and energy via the emission of several $10^{57}$ neutrinos and antineutrinos of all flavors [11, 12]. In the present work we are interested in simple characteristics of the neutrino "light curve" during this quasistatic "Kelvin-Helmholtz cooling phase" such as the total energy emitted in neutrinos, their average energies, and the expected number of events and their time distribution at the Kamiokande II and IMB detectors. In particular, we focus on the variation of these characteristics when the axial-vector neutral-current scattering rates of neutrinos are assumed to be suppressed to a certain degree. As a matter of comparison we also vary other aspects of uncertain input physics such as the equation of state (EOS) and the baryonic mass and temperature profile of the initial protoneutron star model.

### 3.1 Numerical simulations of protoneutron star evolution

The set of initial models used as an input into our evolution simulations was obtained by scaling parameters such as mass and temperature of a protoneutron star model from a core collapse calculation [13] at a stage about half a second after the formation of the SN shock. The EOS of matter around and above nuclear density $\rho_0$ is a finite-temperature extension of a zero-temperature EOS developed by Glendenning [14, 15], which optionally allows us to choose a gas of neutrons, protons, electrons, and muons ("EOS A") or to include additional baryonic states like hyperons and $\Delta$-resonances at densities above $\sim 2\rho_0$ ("EOS B"). For more details about this EOS and the structure of Wilson's post-collapse model, for information about the numerical procedure and the implementation of EOS and neutrino physics, and for a description of the typical evolution of a cooling and neutronizing protoneutron star, see Ref. [12].

The nomenclature of our models was adapted from the paper of Keil and Janka [12]. For example, the letters "SBH" ("**S**mall-EOS **B**-**H**ot") mean a protoneutron star model with a mass smaller than the maximum, stable neutron star mass for EOS B and with an initial central temperature which yields a gravitational mass only little below the baryonic mass, thus ensuring that the model reproduces the marginally bound state of a stellar core shortly after collapse. Correspondingly, "SAH" denotes a model for EOS A. Systematic extensions to these names indicate variations of model properties made in the present work. The sample of investigated models can be grouped in the following way.

1. Models evolved with different neutrino opacities: The degree of the suppression of axial-vector neutral and charged currents is measured by the free parameter $a$ in Eq. (13). The chosen values $a = x.yz$ with $0.00 \leq x.yz \leq 1.00$ are indicated by the extensions "_xyz" in the model names. $a = 0$, i.e. the standard case of the neutrino cross sections, is simply marked by "_0", negative values $-1 \leq a < 0$ (corresponding to enhanced cross sections) are marked by the suffix "m" behind the numbers. The specification "_xyzr" means models where the critical density in Eq. (14) was changed from $3 \times 10^{13}\,\mathrm{g/cm^3}$ to $10^{14}\,\mathrm{g/cm^3}$. When only the neutral currents for neutrino-nucleon interactions are modified, we add an "n" at the end of the names.

2. Models with different masses, initial temperatures, and EOS's: From low to high baryonic mass the models are labelled sequentially as "S0B...", "S1B...", "S2B...", etc. The models of the sequence S#BH_0 were constructed such that the ratio of the initial gravitating mass $M_{\mathrm{G},0}$ to the baryonic mass $M_{\mathrm{B}}$ is constant. Models with a higher initial temperature, i.e. bigger initial gravitational mass, have an asterisk, "*", at the end of their names, even higher temperatures are marked by "**". When EOS A instead of EOS B was used, "B" in the model specification is replaced by an "A".

In Table 1 we present an overview over our sample of models. We list the characteristic parameter combination of each model, i.e. the baryonic mass $M_{\mathrm{B}}$, the initial gravitational mass $M_{\mathrm{G},0}$, the initial baryon number density $n_{\mathrm{c},0}$ at the center of the star, the initial central temperature $T_{\mathrm{c},0}$, and the value of $a$. In addition, the symbols used in the figures are indicated.

Results of the simulation runs are also collected in Table 1. The deleptonization time scale $t_{\mathrm{dl}}$ measures the time until the protoneutron star has reached the deleptonized state, in which the electron lepton number adopts its lowest possible value, corresponding to a chemical potential $\mu_{\nu_e} = 0$ for electron neutrinos. $t_{\mathrm{cl}}$ is the duration of the cooling process by neutrino emission and is defined as the time it takes until the temperatures have dropped below $\sim 3\,\mathrm{MeV}$ in the entire star. $E_{\mathrm{tot}}$ denotes the total energy (redshifted for an observer at infinity) that is radiated away during time $t_{\mathrm{cl}}$, while $t_{90}$ gives the time interval after which 90% of this energy have left the star. The energy emitted in $\overline{\nu}_e$, which may be absorbed on protons in the water Cherenkov detectors of the IMB and Kamiokande II experiments and which are responsible for most of the signals there, is listed in the column "$E_{\overline{\nu}_e}$". We characterize the major differences in the overall behavior and the evolution of different models by these simple parameters. For a description of the typical cooling history of a nascent neutron star we refer the reader to the paper by Keil and Janka [12].

### 3.1.1 Modified neutron star parameters

From the data listed in Table 1 and plotted in Fig. 2 we learn that the time scales $t_{\mathrm{dl}}$, $t_{\mathrm{cl}}$, and $t_{90}$ as well as the energies $E_{\mathrm{tot}}$ and $E_{\overline{\nu}_e}$ increase with growing neutron star mass. The increase of the time scales is moderate and essentially linear for masses $1.3\,\mathrm{M}_\odot \lesssim M \lesssim 1.6\,\mathrm{M}_\odot$, but becomes dramatic for larger masses. The cooling time scale and the time $t_{90}$ vary by more than a factor of 3; $t_{\mathrm{cl}}$ changes from about 28 s for a small star with $M_{\mathrm{B}} \cong 1.30\,\mathrm{M}_\odot$ to nearly 100 s for the $1.75\,\mathrm{M}_\odot$ model, which is very close to the mass limit of stable neutron stars for EOS B, $M_{\mathrm{B}}^{\mathrm{max}} \cong 1.77\,\mathrm{M}_\odot$. The occurrence of hyperons in a large part of the star has a strong influence on the cooling evolution [12]. A more massive protoneutron star loses more gravitational binding energy, the total energy release being roughly proportional to the increase in baryonic mass with a rate of change of about $0.4 \times 10^{53}$ erg per $0.1\,M_\odot$. This dependence becomes less steep towards the high-mass end of our model sample where the production of hyperons consumes a considerable amount of energy and general-relativistic redshift effects gain more importance.

TABLE 1:

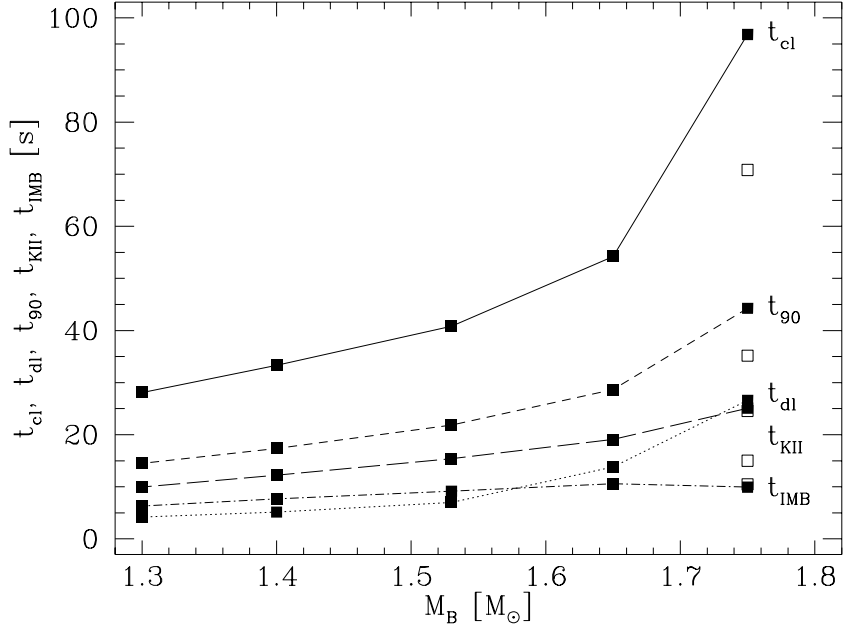

FIG. 2: Cooling time scale $t_{\rm cl}$, deleptonization time scale $t_{\rm dl}$, time interval $t_{90}$ of 90% of the energy loss by neutrino emission, and time intervals $t_{\rm KII}$ and $t_{\rm IMB}$ of 90% of the predicted neutrino counts in the Kamiokande II and IMB detectors, respectively, as functions of the baryonic mass of the protoneutron star. Different plot symbols correspond to the different models listed in Table 1.

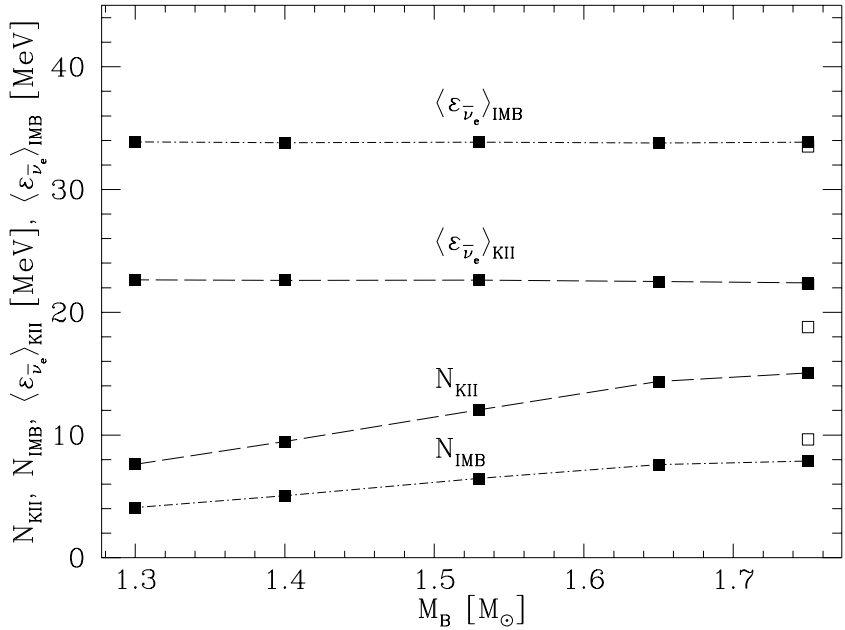

FIG. 3: Predicted numbers of neutrino counts $N_{\rm KII}$ and $N_{\rm IMB}$ in the Kamiokande II and IMB detectors, respectively, and the corresponding average energies $\langle \varepsilon_{\bar\nu_e} \rangle_{\rm KII}$ and $\langle \varepsilon_{\bar\nu_e} \rangle_{\rm IMB}$ of captured electron antineutrinos vs. the baryonic mass of the protoneutron star. Different plot symbols correspond to the different models listed in Table 1.

The effects of redshift and of hyperon production at the high densities in massive protoneutron stars can clearly be seen from a comparison of model S4BH_0 with S4AH_0 which have the same baryonic and the same initial gravitational mass. In model S4AH_0, however, hyperons are absent (EOS A); it releases a larger amount of energy because general-relativistic effects are stronger by about 30% in model S4BH_0, where hyperonization affects as much as the inner $1.2\,M_\odot$. Also, deleptonization and cooling proceed significantly more slowly in model S4BH_0 because the occurrence of hyperons in the central region leads to a more compact core with higher densities and higher temperatures and correspondingly higher neutrino opacities. The peak temperatures in model S4BH_0 reach 65 MeV, in model S4AH_0 they climb up to only 55 MeV. The cold, deleptonized star S4BH_0 has a central baryon density of $1.037\,\mathrm{fm}^{-3}$ and a radius of 10.7 km, whereas the final density of S4AH_0 is $n_\mathrm{c} = 0.519\,\mathrm{fm}^{-3}$ and the radius 12.2 km.

Higher initial temperatures and thus, larger initial gravitating masses (models S2BH_0*, S2BH_0**, and S3BH_0*) lead to an additional energy release of roughly $0.23 \times 10^{53}$ erg per $0.01\,M_\odot$. In all cases, the energy radiated away as $\overline{\nu}_e$'s is about 0.160–0.162 (slightly less than 1/6) of the total energy loss.

### 3.1.2 Modified neutrino opacities

Model S2BH_0 with a baryonic mass of $1.53\,M_\odot$ is used as a reference star for varying the neutrino opacities. Changing the "suppression parameter" $a$ (Sect. 2.2) from $-1$ to 1, we find a monotonic decrease of the time scales $t_\mathrm{cl}$, $t_\mathrm{dl}$, and $t_{90}$. The cooling and deleptonization of the protoneutron star become very fast as $a \to 1$. $t_\mathrm{cl}$ shrinks from a value of 108 s for $a = -1$ (model S2BH_100m) to 41 s for the case of standard opacities, $a = 0$, and drops by another factor of 9 to only 4.5 s for nearly complete suppression of axial-vector weak currents in model S2BH_095 with $a = 0.95$ (Table 1, also Figs. 4 and 5). Interestingly, the relative amount of energy leaving the star as electron antineutrinos decreases monotonically when the stellar matter is assumed to be more transparent to neutrinos.

If the charged-current interactions are left unchanged, the modification of the neutral-current reactions has a somewhat lesser impact than before, in particular the variation of the deleptonization time scale is damped. The overall picture, however, remains unchanged. Replacing the "critical density" in the denominator of Eq. (14) by the value $10^{14}\,\mathrm{g/cm^3}$ has some influence, too, but again the general trends prevail. These results indicate that neutral-current neutrino-nucleon scattering reactions play the dominant role as an opacity source. This is understood because all types of neutrinos ($\nu_e$, $\overline{\nu}_e$, $\nu_\mu$, $\overline{\nu}_\mu$, $\nu_\tau$, and $\overline{\nu}_\tau$) are affected, and because electron neutrino and antineutrino absorptions are strongly blocked due to fermion degeneracy. Moreover, most of the matter of the nascent neutron stars is at densities above $10^{14}\,\mathrm{g/cm^3}$, the surface layers yield only a minor contribution to the optical depth.

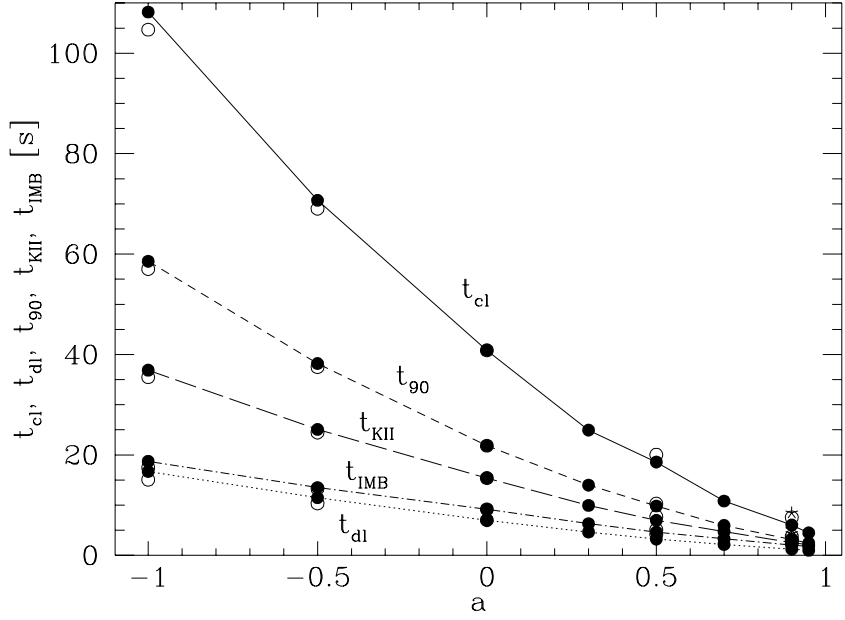

FIG. 4: Cooling time scale $t_{\rm cl}$, deleptonization time scale $t_{\rm dl}$, time interval $t_{90}$ of 90% of the energy loss by neutrino emission, and time intervals $t_{\rm KII}$ and $t_{\rm IMB}$ of 90% of the predicted neutrino counts in the Kamiokande II and IMB detectors, respectively, as functions of the suppression parameter $a$ of axial-vector neutrino-nucleon interactions. The plot symbols correspond to the models listed in Table 1.

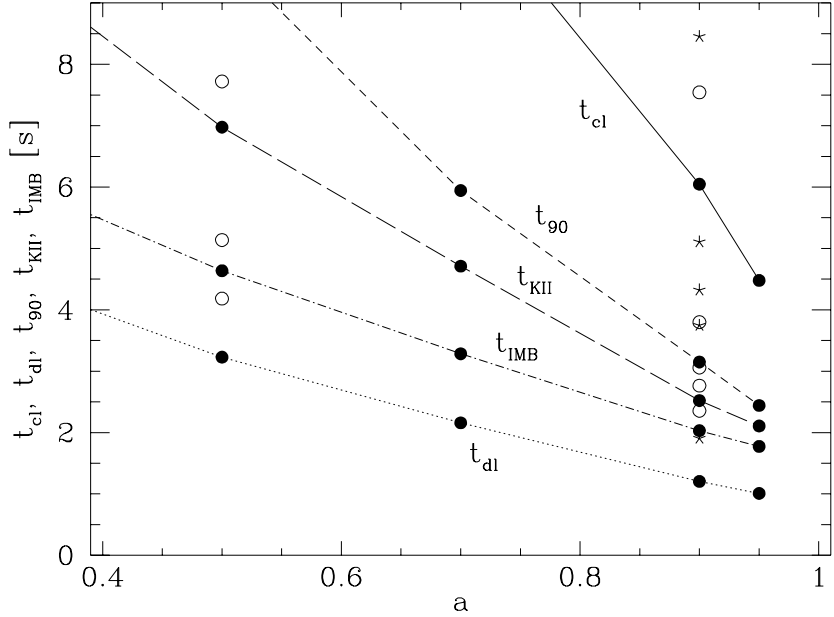

FIG. 5: Enlargement of the region $0.4 \leq a \leq 1$ of Fig. 4.

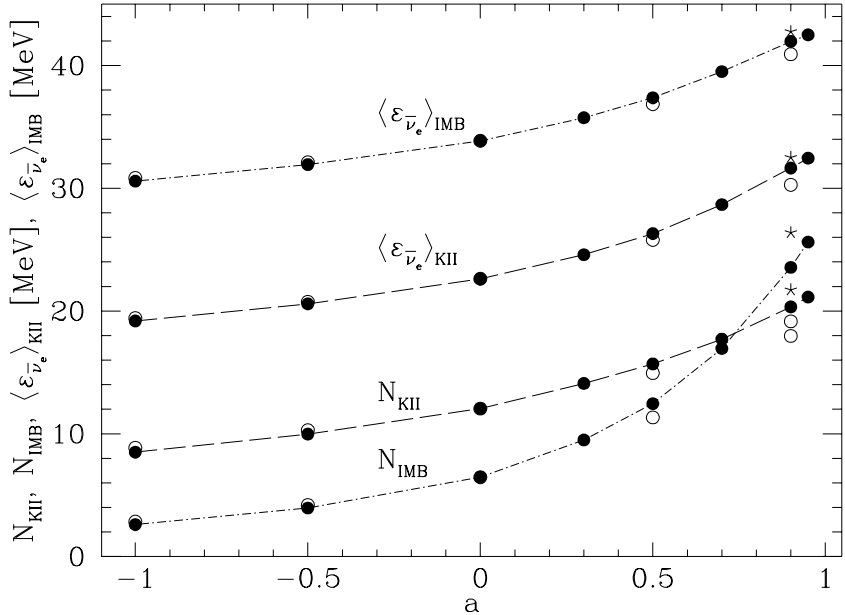

FIG. 6: Predicted numbers of neutrino counts $N_{\rm KII}$ and $N_{\rm IMB}$ in the Kamiokande II and IMB detectors, respectively, and the corresponding average energies $\langle\varepsilon_{\bar{\nu}_e}\rangle_{\rm KII}$ and $\langle\varepsilon_{\bar{\nu}_e}\rangle_{\rm IMB}$ of captured electron antineutrinos vs. the suppression parameter $a$ of axial-vector neutrino-nucleon interactions. The plot symbols correspond to the models listed in Table 1.

## 3.2 Comparison with SN 1987A neutrino signal

In order to judge the degree of parameter variations which is compatible with the SN 1987A neutrino signal, we computed the expected neutrino signals of our models, in particular the signal durations $t_{\rm KII}$ and $t_{\rm IMB}$, the expected event numbers $N_{\rm KII}$ and $N_{\rm IMB}$, and the average energies $\langle\varepsilon_{\bar{\nu}_e}\rangle_{\rm KII}$ and $\langle\varepsilon_{\bar{\nu}_e}\rangle_{\rm IMB}$ of electron antineutrinos absorbed in the Kamiokande II and IMB experiments, respectively. The corresponding data are listed in Table 1. $N_{\rm KII}$ and $N_{\rm IMB}$ represent the total numbers of counts predicted for each model, $t_{\rm KII}$ and $t_{\rm IMB}$ give the time intervals in which 90% of them occur. The source was assumed at a distance of 50 kpc, the approximate distance of the Large Magellanic Cloud. Details of the procedure to evaluate our models to obtain these data can be found in the appendix of the work by Keil and Janka [12].

### 3.2.1 Modified neutron star parameters

The average energies of the neutrinos measured in both detectors reveal an amazing stability against differences of the masses of our models (see Table 1 and Fig. 3). This indicates some form of self-regulation of the position of the neutrino sphere, which is the layer below the surface of the star where neutrinos decouple from the stellar background. Due to the higher core temperatures of more massive stars the neutrinos are more energetic. However, the neutrino opacity of the stellar matter also increases because the interaction rates of neutrinos depend on the particle energies and thus on the temperature. As a result the position of the neutrino sphere moves further out into the cooler regions closer to the surface of the star.

The number of events and signal duration are increasing functions of the neutron star mass (Table 1, Figs. 2 and 3). Deviations from the essentially linear dependence become visible again for our most massive model S4BH_0 with $M_B = 1.75 \, M_\odot$. The signal duration $t_{\rm KII}$ reflects the steep growth of the deleptonization and cooling times for the heavy stars (Fig. 2). In contrast, $t_{\rm IMB}$ tends to decrease (Fig. 2), and $N_{\rm KII}$ and $N_{\rm IMB}$ are nearly inert to the larger energy loss of model S4BH_0 compared with model S3BH_0 ($M_B = 1.65 \, M_\odot$), see Fig. 3 and Table 1. This behavior is caused by the strong gravitational redshift of the massive model S4BH_0 which is very compact because hyperonization softens the EOS in a large part of the star. In particular, the IMB experiment with its high energy threshold (about 20 MeV) is sensitive to the corresponding suppression of the high energies in the neutrino spectrum. A comparison between models S4BH_0 and S4AH_0 confirms this interpretation and again demonstrates the considerable effects that result from disregarding hyperons in the much stiffer EOS A.

Our initial models were constructed "by hand" rather than obtained from a self-consistent collapse calculation. This entails an uncertainty which may be estimated from the changes of the calculated detector responses for models with modified initial temperatures (gravitating masses), i.e. models S2BH_0, S2BH_0*, and S2BH_0**, and models S3BH_0 and S3BH_0*. The signal durations are completely unaffected while the number of counts scatters by 2–4 units in the investigated cases.

Provided "standard" neutrino opacities and EOS B yield a reasonable description of the physics at the extreme conditions in SN cores, our sample of cooling simulations with modified neutron star masses allows for the conclusion that the SN 1987A neutrino pulse most likely indicates a protoneutron star with a baryonic mass between 1.5 and 1.6 $M_\odot$. We caution, however, that in the present study we did not intend to find the best-fitting model for the neutrino source in SN 1987A.

### 3.2.2 Modified neutrino opacities

Reduced or enhanced neutrino opacities cause clear changes of the neutrino signature in the Kamiokande II and IMB experiments. The predicted signal durations follow the trends of $t_{\rm dl}$, $t_{\rm cl}$, and $t_{90}$. For the very opaque case ($a = -1$, model S2BH_100m) Kamiokande II and IMB measure events over periods of 37 s and 19 s, respectively (Table 1 and Fig. 4), while for a nearly complete suppression of the axial-vector contributions to neutrino-nucleon interactions ($a = 0.95$, model S2BH_095) the signal durations drop to about 2 s in both detectors (Fig. 5).

Interestingly, the expected number of counts and also the mean energies of neutrinos captured in the detectors show the inverse behavior. When the value of $a$ is changed from $a = -1$ to 0.95, $N_{\rm KII}$ increases from 9 to 21, and $N_{\rm IMB}$ from 3 to 26. This strong sensitivity can be understood by the spectral characteristics of the emitted neutrinos. Neutrinos decouple from the stellar medium at an optical depth of about unity. For a low neutrino opacity of the neutron star matter they leave the star from deeper inside, i.e. from layers of higher densities and temperatures. Therefore, these neutrinos are more energetic as is evident from the monotonic rise of $\langle \varepsilon_{\bar\nu_e} \rangle_{\rm KII}$ and $\langle \varepsilon_{\bar\nu_e} \rangle_{\rm IMB}$ with higher values of the suppression factor $a$ (Fig. 6).

Modifying only neutral-current interactions while keeping the charged currents fixed (models marked with an "n" at the end of their names) leads to changes of the predicted neutrino signals relative to our standard case S2BH_0 which are only slightly less dramatic. For model S2BH_090r with a higher value of the critical density, we find, in contrast, an even stronger increase of the mean energies of the captured neutrinos and, correspondingly, of the number of detector counts than for model S2BH_090 or model S2BH_090n. This is caused by the stronger heating of the outer regions of the star where the cooling of the low-opacity inner part is slowed down by the less transparent surface layer.

The opposite trends of the different signal parameters (decreasing cooling and deleptonization time scales, increasing count rates and mean neutrino energies) allow us to exclude a certain range of $a$ as being highly unlikely. The extremely short signal durations and the very large numbers of energetic neutrino events predicted for models S2BH_090, S2BH_095, S2BH_090n, and S2BH_090r seem to rule out a complete or nearly complete suppression of axial-vector current interactions[2]. Neither statistical fluctuations of the data nor theoretical uncertainties like the EOS of supranuclear matter or the baryonic and gravitational mass of the protoneutron star seem to be able to account for the signal characteristics that obtain in the limit $a \to 1$.

### 3.2.3 Modified neutrino spectra

The total number of events observed in the IMB detector is rather sensitive to the high-energy tail of the $\overline{\nu}_e$-flux because of the relatively high energy threshold of IMB. In the protoneutron star cooling calculations presented here, neutrino transport was treated by an equilibrium-diffusion method [12], implying that the spectral $\overline{\nu}_e$-distribution is essentially a Fermi-Dirac function with a degeneracy parameter $\eta = 0$. A detailed Monte Carlo treatment of neutrino transport [16] reveals, however, that the $\overline{\nu}_e$-spectrum is probably much better approximated by a Fermi-Dirac function with $\eta = 2$–3. If the average $\overline{\nu}_e$-energy is kept fixed this implies that both the low- and high-energy parts of the spectrum are depleted relative to intermediate energies, i.e., the spectral distribution is "pinched". This effect would cause a substantial depletion of the observable number of events at IMB.

In order to study the impact of this effect quantitatively we took the $\overline{\nu}_e$-flux and average $\overline{\nu}_e$-energies $\langle \varepsilon_{\bar\nu_e} \rangle$ of our reference case (model S2BH_0) and calculated the expected neutrino counts, the average $e^+$-energies in the Kamiokande II and IMB detectors, and the expected signal durations as functions of $\eta$. We stress, again, that $\langle \varepsilon_{\bar\nu_e} \rangle$ was held fixed at a given time, not the spectral temperature, which was adjusted for a given $\eta$ to yield the same $\langle \varepsilon_{\bar\nu_e} \rangle$. In Fig. 7 we show the relative change of $N_{\mathrm{KII}}$, $N_{\mathrm{IMB}}$, $\langle \varepsilon_{e^+} \rangle_{\mathrm{KII}}$, $\langle \varepsilon_{e^+} \rangle_{\mathrm{IMB}}$, $t_{\mathrm{KII}}$, and $t_{\mathrm{IMB}}$. Only $N_{\mathrm{IMB}}$ responds sensitively to $\eta$. For the standard range $\eta = 2 - 3$ one obtains a reduction between 30% and 50%.

The uncertainty introduced by the unknown amount of pinching, the possibility that the neutrino spectra may have been modified by oscillations (e.g. Ref. [17]), and the small integrated number of events in each detector together reveal that the total numbers of events, and especially $N_{\rm IMB}$, are rather poor measures to learn much about the properties of the neutrino signal. The most sensitive observables remain the durations of the signals registered at the two detectors.

---

[2]The Kamiokande II signal consists of eight events within about 2 s and another three events between about 9 and 12 s after the first one. Our conclusions are based on taking the late-time events as part of the neutrino signal, i.e., the long time gap of about 7 s at Kamiokande is interpreted as a statistical fluctuation.

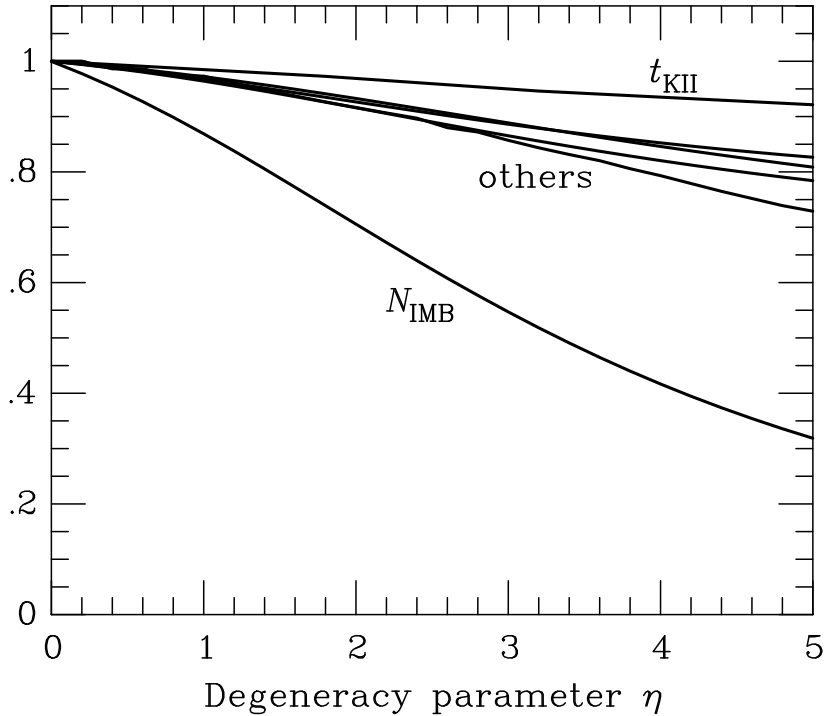

FIG. 7: Impact of the "spectral pinching" on the total numbers of events in the Kamiokande II and IMB detectors ($N_{\rm KII}$, $N_{\rm IMB}$), on the average energies of the observed positrons from the reaction $\bar{\nu}_e + p \to n + e^+$, and on the expected signal durations $t_{\rm KII}$ and $t_{\rm IMB}$.

## 4  Discussion and conclusions

Recently, Raffelt and Seckel [6] conjectured that the axial-vector part of weak neutral currents — possibly also of weak charged currents — in neutrino-nucleon interactions should be (partially) suppressed at the high densities present in SN cores. This effect would be a consequence of fast fluctuations of the nucleon spins due to frequent collisions. In the present work we investigated the implications of this possibility in the astrophysical scenario of neutron star formation. For that purpose we performed numerical simulations of the neutrino-driven cooling and deleptonization of newly formed neutron stars and determined the changes associated with a reduction of the neutrino cross sections. We computed the expected neutrino signals for the Kamiokande II and IMB detectors. Consistency between these theoretical predictions and the SN 1987A neutrino observations can be used to obtain limits on the acceptable range of neutrino opacities.

Our simulations showed that the duration of neutrino emission from the nascent neutron star and the corresponding signal durations at the detectors are drastically reduced in the limit of a (nearly) complete suppression of the axial-vector current interactions. Interestingly, due to the increased transparency of neutron star matter for neutrinos this is accompanied by a strong rise of the number of neutrino counts and, in particular, of the average energies of the absorbed electron antineutrinos. These diverging trends lead to a clear discrepancy of the characteristics of the theoretical neutrino signals with the observed ones.

Testing the sensitivity of the signal against variations of unknown properties of the nascent neutron star (mass of the protoneutron star, initial temperature profile, EOS at supranuclear densities) we found that the associated uncertainties cannot compensate the changes from extremely suppressed neutrino opacities. In particular, it seems impossible to construct a scenario where both of the diverging trends are simultaneously compensated. For example, a decrease of the signal time due to reduced neutrino opacities of the protoneutron star could be moderated by a higher mass of the star. However, in that case the number of events will rise even more. Therefore, our numerical results demonstrate that a strong suppression of the axial-vector part of neutrino-nucleon interactions can be firmly excluded. Being more quantitative, it appears very unlikely that suppression factors of $a \gtrsim 0.5$ do occur.

Concerning some hypothetical mechanism that might lead to an amplification of neutrino-nucleon interactions at high densities we are not able to draw equally definite conclusions. Even though the signal durations become appreciably longer when the parameter $a$ gets negative, this could be compensated, at least in part, by the faster cooling of a smaller protoneutron star. The mean energies of the captured neutrinos do not display a similar sensitivity for $a < 0$ and thus cannot serve to put strong limits on an acceptable enhancement of neutrino opacities. Nevertheless, our results indicate that a slight decrease of $\langle \varepsilon_{\bar{\nu}_e} \rangle_{\mathrm{IMB}}$ is already sufficient to lead to a considerable decrease of the number of neutrino counts in the IMB experiment. This adds to the already suppressed number of neutrino events from a less massive neutron star. Therefore, an enhancement of the neutrino opacities by more than 50% can be considered as quite improbable and by a factor of 2 or more seems to be excluded. However, we stress that the count numbers in the detectors, in particular in the IMB detector, are rather poor measures of the properties of the neutrino emitting star. On the one hand they are sensitive to the uncertain amount of spectral pinching (Sect. 3.2.3), on the other hand they turned out to change noticeably with the temperature of the star (Sect. 3.2.1) or by additional neutrino emission due to mass accretion onto the newly formed neutron star [18].

Therefore, the observed neutrino signal indicates that the neutrino opacities cannot be too different from what a naive cross section calculation for $\nu + N \to N + \nu$ yields which, in turn, is based on a spin structure function of the form $S_\sigma(\omega) = 2\pi\delta(\omega)$. This does not imply, however, that a naive perturbative calculation of all microscopic reaction rates involving the axial-vector current was essentially appropriate. A naive calculation of the bremsstrahlung process $N + N \to N + N + \bar{\nu} + \nu$ or $N + N \to N + N + a$ (axion $a$) entails that $S_\sigma(\omega)$ should be a broad function of width $\Gamma_\sigma \gg T$ where the "spin fluctuation rate" $\Gamma_\sigma$ was defined in Eq. (6). Our findings indicate that $S_\sigma(\omega)$ is a function with a width not larger than a few $T$; this conclusion depends on the normalization condition of $S_\sigma$, i.e., the assumption of uncorrelated spin motions. Put another way, our findings appear to imply that the spin fluctuation rate in a nuclear medium is of order $T$ and not as large as given by a naive perturbative estimate.

Even though the neutrino opacity appears to be similar to what a calculation based on free nucleons yields and what was used in practical applications in the past, the axial-vector structure function is, no doubt, very broad rather than a $\delta$-function. Therefore, in a single collision the energy of the neutrino typically changes by a large amount. This is in contrast to the standard picture where the neutral-current collision on "heavy" nucleons (small recoil) conserves the neutrino energy. According to Eq. (6) the width of $S_\sigma(\omega)$ is around $3T$ already at 10% nuclear density so that the picture of energy-conserving neutral-current collisions breaks down at relatively low densities, whether or not the structure function becomes even broader at higher densities. (Our study suggests that it does not become much broader.) We have not addressed this issue in our present work, even though it may have an impact on the predicted neutrino spectrum and notably on the "pinching effect" found in the Monte Carlo simulations of Janka and Hillebrandt [16].

The bremsstrahlung emission of neutrino pairs or axions is given by an integral of the type Eq. (4); it involves a factor $e^{-\omega/T}$ which is included in $S_\sigma(\omega)$ for positive energy transfers. For a normalized $S_\sigma(\omega)$ this integral is maximal if $S_\sigma(\omega)$ is a broad function with a width of a few $T$. As our results seem to indicate that $S_\sigma(\omega)$ does not grow much broader than a few $T$ one is led to conclude that the bremsstrahlung emission rate saturates at its maximum possible value which is reached at around 10% nuclear density, without dropping at yet larger densities. Granting this, the axion emission rate from a SN core is most realistically estimated by its value at the given temperatures and about 10% nuclear density. For typical axion models this means that $m_a \gtrsim 10^{-2}\,\text{eV}$ is excluded, i.e., previous SN 1987A axion bounds [5] are not diminished as much as one would have feared on the basis of the extreme suppression scenario discussed in Sect. 2.

While the SN 1987A neutrino signal indicates that the neutrino opacities are not too far from their "standard" values, their actual magnitudes have not been computed on the basis of first principles and so, they essentially remain adjustable parameters to account for a SN neutrino light curve. Because the opacities are dominated by the axial-vector neutral-current interactions with the nuclear medium one can estimate a plausible functional form of the spin-structure function $S_\sigma(\omega)$ which, in turn, governs quantities such as the axion emission rate. Therefore, even though the axion emission rate can not be calculated by simple perturbative means one can estimate its magnitude from the "observed" neutrino opacities. As a result, a SN neutrino signal remains a useful tool to constrain particle physics phenomena based on the axial-vector coupling such as the axion-nucleon coupling.

**Acknowledgements.** We are very grateful to N.K. Glendenning for providing us with the tables of his equations of state. We also want to thank J.R. Wilson for the data of the protoneutron star from one of his SN simulations, which we used to construct some of the initial models for our investigations. Furthermore we acknowledge the Master's research done by T. Hecht [19], from which the numerical implementation of the EOS's was taken. G. R. acknowledges the hospitality of the Center for Particle Astrophysics during part of this research.

TABLE 1: Compilation of protoneutron star models with their characteristic parameters, the baryonic mass $M_{\rm B}$, and the initial values of gravitational mass $M_{\rm G,0}$, central baryon density $n_{\rm c,0}$, and central temperature $T_{\rm c,0}$. In addition, the suppression factor $a$ of axial-vector contributions to neutrino-nucleon interactions and the plot symbol are given for each model. The neutrino-driven evolution of the protoneutron stars is described by the time scales of the deleptonization process, $t_{\rm dl}$, and of the cooling, $t_{\rm cl}$. $t_{90}$ is defined as the time interval after which 90% of the total energy emitted in neutrinos, $E_{\rm tot}$, have left the star. $E_{\rm tot}$ is the energy measured by an observer at infinity. The fraction of this energy which is released as electron antineutrinos is given by $E_{\bar{\nu}_e}$. As for the predicted neutrino signals in the Kamiokande II and IMB detectors, we list the integrated numbers of events in the detectors, $N_{\rm KII}$ and $N_{\rm IMB}$, the time intervals $t_{\rm KII}$ and $t_{\rm IMB}$ in which 90% of these events occur, and the average energies of the captured electron antineutrinos, $\langle\varepsilon_{\bar{\nu}_e}\rangle_{\rm KII}$ and $\langle\varepsilon_{\bar{\nu}_e}\rangle_{\rm IMB}$, respectively.

| Model | | $M_{\rm B}$ (M$_\odot$) | $M_{\rm G,0}$ (M$_\odot$) | $n_{\rm c,0}$ (fm$^{-3}$) | $T_{\rm c,0}$ (MeV) | $a$ | $t_{\rm dl}$ (s) | $t_{\rm cl}$ (s) | $t_{90}$ (s) | $E_{\rm tot}$ ($10^{53}$ erg) | $E_{\bar{\nu}_e}$ ($10^{52}$ erg) | $N_{\rm KII}$ | $t_{\rm KII}$ (s) | $\langle\varepsilon_{\bar{\nu}_e}\rangle_{\rm KII}$ (MeV) | $N_{\rm IMB}$ | $t_{\rm IMB}$ (s) | $\langle\varepsilon_{\bar{\nu}_e}\rangle_{\rm IMB}$ (MeV) |
|---|---|---|---|---|---|---|---|---|---|---|---|---|---|---|---|---|---|
| S0BH_0 | ■ | 1.30 | 1.26 | 0.400 | 16.6 | 0.00 | 4.2 | 28.1 | 14.5 | 1.63 | 2.62 | 7.6 | 9.9 | 22.6 | 4.1 | 6.3 | 33.8 |
| S1BH_0 | ■ | 1.40 | 1.37 | 0.435 | 20.2 | 0.00 | 5.2 | 33.3 | 17.4 | 2.03 | 3.25 | 9.5 | 12.3 | 22.6 | 5.1 | 7.7 | 33.8 |
| S2BH_0 | ■ | 1.53 | 1.49 | 0.490 | 24.0 | 0.00 | 7.0 | 40.8 | 21.8 | 2.56 | 4.14 | 12.1 | 15.0 | 22.6 | 6.5 | 9.2 | 33.8 |
| S2BH_0* | | 1.53 | 1.51 | 0.480 | 27.3 | 0.00 | 7.4 | 43.0 | 22.2 | 3.00 | 4.83 | 14.7 | 15.3 | 23.2 | 8.4 | 9.3 | 34.4 |
| S2BH_0** | | 1.53 | 1.52 | 0.470 | 28.4 | 0.00 | 7.3 | 43.7 | 21.9 | 3.23 | 5.21 | 16.3 | 15.0 | 23.8 | 9.8 | 9.0 | 34.8 |
| S3BH_0 | ■ | 1.65 | 1.61 | 0.560 | 28.1 | 0.00 | 13.9 | 54.2 | 28.7 | 3.11 | 5.04 | 14.4 | 19.1 | 22.5 | 7.6 | 10.6 | 33.7 |
| S3BH_0* | | 1.65 | 1.63 | 0.551 | 30.7 | 0.00 | 14.9 | 55.4 | 28.7 | 3.57 | 5.76 | 17.2 | 19.2 | 23.1 | 9.7 | 10.6 | 34.3 |
| S4BH_0 | ■ | 1.75 | 1.71 | 0.637 | 31.9 | 0.00 | 26.6 | 96.8 | 44.3 | 3.43 | 5.55 | 15.1 | 25.1 | 22.4 | 7.9 | 9.9 | 33.8 |
| S4AH_0 | □ | 1.75 | 1.71 | 0.654 | 31.9 | 0.00 | 10.4 | 70.8 | 35.2 | 4.11 | 6.65 | 18.8 | 24.6 | 22.3 | 9.7 | 15.0 | 33.5 |
| S2BH_100m | ● | 1.53 | 1.49 | 0.490 | 24.0 | −1.00 | 17.0 | 108.2 | 58.6 | 2.58 | 4.24 | 8.5 | 36.9 | 19.2 | 2.6 | 18.7 | 30.7 |
| S2BH_050m | ● | 1.53 | 1.49 | 0.490 | 24.0 | −0.50 | 11.5 | 70.7 | 38.2 | 2.57 | 4.19 | 10.0 | 25.1 | 20.6 | 3.9 | 13.5 | 31.9 |
| S2BH_0 | ● | 1.53 | 1.49 | 0.490 | 24.0 | 0.00 | 7.0 | 40.8 | 21.8 | 2.56 | 4.14 | 12.1 | 15.0 | 22.6 | 6.5 | 9.2 | 33.8 |
| S2BH_030 | ● | 1.53 | 1.49 | 0.490 | 24.0 | 0.30 | 4.7 | 24.9 | 14.0 | 2.55 | 4.16 | 14.1 | 9.9 | 24.5 | 9.5 | 6.3 | 35.8 |
| S2BH_050 | ● | 1.53 | 1.49 | 0.490 | 24.0 | 0.50 | 3.4 | 18.6 | 9.8 | 2.55 | 4.15 | 15.7 | 7.0 | 26.3 | 12.5 | 4.6 | 37.4 |
| S2BH_070 | ● | 1.53 | 1.49 | 0.490 | 24.0 | 0.70 | 2.2 | 10.8 | 5.9 | 2.54 | 4.09 | 17.7 | 4.7 | 28.7 | 17.0 | 3.3 | 39.5 |
| S2BH_090 | ● | 1.53 | 1.49 | 0.490 | 24.0 | 0.90 | 1.2 | 6.1 | 3.2 | 2.56 | 4.06 | 20.3 | 2.5 | 31.6 | 23.7 | 2.0 | 41.9 |
| S2BH_095 | ● | 1.53 | 1.49 | 0.490 | 24.0 | 0.95 | 1.0 | 4.5 | 2.4 | 2.56 | 4.05 | 21.1 | 2.1 | 32.5 | 25.6 | 1.8 | 42.5 |
| S2BH_100mn | ○ | 1.53 | 1.49 | 0.490 | 24.0 | −1.00n | 15.1 | 104.7 | 57.0 | 2.59 | 4.31 | 8.9 | 35.5 | 19.4 | 2.8 | 17.5 | 30.8 |
| S2BH_050mn | ○ | 1.53 | 1.49 | 0.490 | 24.0 | −0.50n | 10.5 | 69.0 | 37.5 | 2.58 | 4.25 | 10.3 | 24.4 | 20.7 | 4.2 | 13.3 | 32.1 |
| S2BH_0 | ○ | 1.53 | 1.49 | 0.490 | 24.0 | 0.00 | 7.0 | 40.8 | 21.8 | 2.56 | 4.14 | 12.1 | 15.0 | 22.6 | 6.5 | 9.2 | 33.8 |
| S2BH_050n | ○ | 1.53 | 1.49 | 0.490 | 24.0 | 0.50n | 4.3 | 20.0 | 10.3 | 2.53 | 4.09 | 15.0 | 7.7 | 25.8 | 11.3 | 5.1 | 36.9 |
| S2BH_090n | ○ | 1.53 | 1.49 | 0.490 | 24.0 | 0.90n | 2.8 | 7.5 | 3.8 | 2.46 | 3.85 | 18.0 | 3.1 | 30.3 | 19.0 | 2.4 | 40.9 |
| S2BH_090r | ∗ | 1.53 | 1.49 | 0.490 | 24.0 | 0.90 | 2.0 | 8.5 | 5.1 | 2.58 | 4.19 | 21.7 | 4.3 | 32.5 | 26.4 | 3.7 | 42.8 |
| SN1987A | — | — | — | — | — | — | — | — | — | — | — | 3–6 | 11 | 12.4 | 17.4 | 8 | 5.6 | 33.9 |